\documentclass[lettersize,journal]{IEEEtran}
\usepackage{amsmath,amsfonts}
\usepackage{algorithmic}
\usepackage{algorithm}
\usepackage{array}
\usepackage[caption=false,font=normalsize,labelfont=sf,textfont=sf]{subfig}
\usepackage{textcomp}
\usepackage{stfloats}
\usepackage{url}
\usepackage{verbatim}
\usepackage{graphicx}
\usepackage{cite}
\usepackage{siunitx}
\usepackage{booktabs}
\usepackage[colorlinks = true,
            linkcolor = blue,
            urlcolor  = blue,
            citecolor =  green,
            anchorcolor = blue]{hyperref}
\usepackage{xcolor}

\begin{document}

\newcommand{\vModelName}{\textit{AudioLDM 2}}

\title{AudioLDM 2: Learning Holistic Audio Generation with Self-supervised Pretraining}

\author{Haohe Liu, Yi Yuan, Xubo Liu, Xinhao Mei, Qiuqiang Kong, \\
Qiao Tian, Yuping Wang, Wenwu Wang, Yuxuan Wang, Mark D. Plumbley \\ 
\thanks{Haohe Liu, Yi Yuan, Xubo Liu, Xinhao Mei, Wenwu Wang, and Mark D. Plumbley are with the Centre for Vision, Speech and Signal Processing (CVSSP), University of Surrey, Guilford, UK. Email: \{haohe.liu, yi.yuan, xubo.liu, x.mei, w.wang, m.plumbley\}@surrey.ac.uk.}
\thanks{Qiuqiang Kong is with the Department of Electronic Engineering, Chinese University of Hong Kong, Hong Kong, China. Email: qqkong@ee.cuhk.edu.hk}
\thanks{Qiao Tian, Yuping Wang and Yuxuan Wang: are with the Speech, Audio \& Music Intelligence (SAMI) Group, ByteDance Inc. Email: \{tianqiao.wave, kongqiuqiang, wangyuping, wangyuxuan.11\}@bytedance.com.}}

\maketitle

\begin{abstract}
Although audio generation shares commonalities across different types of audio, such as speech, music, and sound effects, designing models for each type requires careful consideration of specific objectives and biases that can significantly differ from those of other types. To bring us closer to a unified perspective of audio generation, this paper proposes a \textcolor{black}{holistic} framework that utilizes the same learning method for speech, music, and sound effect generation. Our framework \textcolor{black}{utilizes} a general representation of audio, called ``language of audio''~(LOA). Any audio can be translated into LOA based on AudioMAE, a self-supervised pre-trained representation learning model. In the generation process, we translate \textcolor{black}{other} modalities into LOA by using a GPT-2 model, and we perform self-supervised audio generation learning with a latent diffusion model conditioned on the LOA \textcolor{black}{of audio in our training set}. The proposed framework naturally brings advantages such as reusable self-supervised pretrained latent diffusion models. Experiments on the major benchmarks of text-to-audio, text-to-music, and text-to-speech~\textcolor{black}{with three~\vModelName~variants} demonstrate competitive performance of the~\vModelName~\textcolor{black}{framework} against previous approaches. Our code, pretrained model, and demo are available at \url{https://audioldm.github.io/audioldm2}.
\end{abstract}


\section{Introduction}

\IEEEPARstart{A}RTIFICIAL intelligence generated content~(AIGC) refers to any digital content such as images, videos, text, or audio that has been fully or partially created by an AI system without human involvement in the creative process~\cite{cao2023comprehensive}. Of particular interest is the ability of AI to produce audio content based on text, phonemes, or images~\cite{tan2022naturalspeech, kreuk2022audiogen, liu2023audioldm}. AI-based audio generation has a wide potential in applications including synthesizing human or artificial voices for digital assistants~\cite{zhang2023speechgpt}, generating sound effects and background music for movies, and games~\cite{riffusion}, and automating the production of podcasts and audiobooks~\cite{liu2023wavjourney}. 

AI-based audio generation is often undertaken in separate sub-domains, such as the generation of speech~\cite{tan2022naturalspeech}, music~\cite{agostinelli2023musiclm}, sound effects~\cite{liu2023audioldm}, and specific types of sounds such as footsteps and violin sounds~\cite{bresin2010expressive,engel2020ddsp}. To address the specific challenges in each sub-domain, most previous works design task-specific inductive biases, which are predefined constraints that guide the learning process to a specific problem space.
For example, pitch and duration predictors are often used in speech synthesis to model the prosody of speech~\cite{Fastspeech2, tan2022naturalspeech}, while MIDI representation~\cite{herremans2016morpheus} and domain-specific pre-trained modules are often used in music generation~\cite{lam2023efficient-melody, agostinelli2023musiclm}.

Despite significant progress being made in developing specialized models for specific sub-domains of audio generation, the limitations of such specialization restrict the broader application of audio-generation models in complex auditory scenarios. 
\textcolor{black}{Although there are models that can generate various types of audio, such as AudioLDM~\cite{liu2023audioldm}, the speech they generate is still not intelligible.}
Whether a unified approach can be developed to generate various types of audio signals\textcolor{black}{, including intelligible speech,} remains unanswered.
Different types of sound can occur simultaneously in real-world cases, such as in movie scenes, requiring a more general approach to modelling audio generation. 
While there are works that address audio generation in a general domain, they mostly focus on generating audio with correct audio events with limited attention to detail. For example, previous text-to-audio generation research tends to generate unintelligible speech~\cite{yang2022diffsound,liu2023audioldm,huang2023make-an-audio}.
Moreover, while inductive biases have been useful in addressing the challenges of specific sub-domains, conclusions about a specific design drawn from one domain may not necessarily transfer to another. 
Recent advancements in addressing problems from a unified perspective have yielded substantial progress~\cite{liu2021voicefixer, baevski2022data2vec, girdhar2023imagebind, kong2023universal}. This trend highlights the potential of constructing a unified audio generation framework.

This paper presents a novel and versatile framework, called \vModelName, that can generate audio with flexible conditions, without the need for domain-specific inductive bias. 
\textcolor{black}{The core idea is to introduce a sequence of vectors that represent the semantic information of an audio clip, which we will refer to as the ``language of audio''~(LOA).}
This approach allows us to translate human-understandable information into LOA and synthesize audio representation conditioned on LOA. This idea is similar to the use of onomatopoeia in~\cite{okamoto2022onoma} to describe environmental sounds. However, although onomatopoeia can effectively mimic certain sounds like animal noises or simple actions (e.g., ``splash'' for water), it can not encompass the full range of audio nuances. 
In theory, the ``language of audio'' should be able to represent both fine-grained acoustic information~(e.g., ``what does the speaker say'') and coarse-grained semantic information~(e.g., ``what is that sound''). Considering these requirements, we propose to leverage the features extracted by an audio masked autoencoder~(AudioMAE)~\cite{xu2022masked}, an audio-generative self-supervised pretraining framework. An AudioMAE is pre-trained on diverse audio content, and its dual generative and reconstructive pre-training approach makes it potentially a strong option for representing audio in generative tasks.

Specifically, we utilize a GPT-2 language model~\cite{radford2019language} to translate conditioning information into the AudioMAE features. 
We then use a latent diffusion model~\cite{rombach2022high-stablediffusion} to synthesize audio based on the AudioMAE features. The latent diffusion model can be optimized in a self-supervised manner, allowing for pre-training with large-scale unlabelled audio data. Our language-modelling approach with GPT-2 enables us to leverage recent advancements in language models~\cite{zhao2023survey}, while alleviating challenges such as high inference computation costs and error accumulation that appeared in previous audio autoregressive models~\cite{zeghidour2021soundstream, agostinelli2023musiclm}. The improvement is largely attributed to the shorter length of the LOA sequence. 
The continuous nature of LOA also potentially provides a richer representation power than the discrete tokens used in previous models ~\cite{lam2023efficient-melody, borsos2023audiolm, agostinelli2023musiclm}.

Our experimental results demonstrate that \vModelName~achieves \textcolor{black}{competitive} performance on text-to-audio~(TTA), and text-to-music~(TTM) generation tasks, when evaluated on AudioCaps~\cite{kim2019audiocaps} and MusicCaps~\cite{agostinelli2023musiclm}, respectively. On text-to-speech~(TTS) generation tasks, \vModelName~achieves performance comparable with the SoTA by significantly outperforming a strong baseline FastSpeech2~\cite{Fastspeech2}. 
In comparison to the original AudioLDM~\cite{liu2023audioldm}, \textcolor{black}{~\vModelName~contains a latent diffusion model that can be pretrained in a self-supervised manner}, and enjoy the benefit of auto-regressive modeling of LOA with GPT-2 model. Besides, while retaining the same ability, \vModelName~shows substantial advancements over AudioLDM in quality, versatility, and capacity to generate speech with intelligible content. 
Overall, our contributions are as follows:

\begin{list}{\labelitemi}{\leftmargin=1em}
    \item We propose a novel and versatile audio generation model that is capable of performing conditional generation of audio, music, and intelligible speech. 
    \item The proposed method is based on a universal representation of audio, which enables large-scale self-supervised pretraining of the core latent diffusion model without audio annotation and helps to combine the advantages of both the auto-regressive and the latent diffusion model. 
    \item Our experiments shows \textcolor{black}{three variants of} \vModelName~achieves \textcolor{black}{performance that match current state-of-the-art (SoTA) in text-to-audio, text-to-music, and text-to-speech generation on AudioCaps~\cite{kim2019audiocaps}, MusicCaps~\cite{agostinelli2023musiclm}, and LJSpeech~\cite{ljspeech17} evaluation set, respectively.}
\end{list}

\section{Related Work}

\subsection{Conditional Audio Generation}

\noindent
Audio generation is an emerging topic that focuses on modelling the generation of general audio, including recent models such as AudioGen~\cite{kreuk2022audiogen}, AudioLDM~\cite{liu2023audioldm}, and Make-an-Audio~\cite{huang2023make-an-audio}. AudioGen treats audio generation as a conditional language modelling task, while the other two works approach this task by latent diffusion. Studies on image-to-audio and video-to-audio generation, such as~Im2Wav~\cite{sheffer2023hear} and SpecVQGAN~\cite{iashin2021taming-specvqgan}, are also areas of interest to researchers. Additionally, there are audio generation approaches that do not rely on conditioning, such as AudioLM~\cite{borsos2023audiolm}, which performs audio language modelling based on a neural codec. Even though audio generation usually includes the topic of speech generation, previous works on text-to-audio generation tend to generate unintelligible speech~\cite{yang2022diffsound,liu2023audioldm,kreuk2022audiogen,huang2023make-an-audio}.

The field of audio generation encompasses sub-domains such as text-to-speech (TTS) and text-to-music (TTM). The former focuses on generating speech signals from transcriptions, while the latter involves creating a music clip from a textual description. Cutting-edge TTS models like FastSpeech2~\cite{Fastspeech2}, GradTTS~\cite{popov2021gradtts}, and NaturalSpeech~\cite{tan2022naturalspeech} have made significant strides, producing speech of such high quality that it is nearly indistinguishable from human speech. Various techniques have been introduced to address speech generation in TTS, such as the monotonic alignment algorithm~\cite{kim2020glowtts}, which aligns phoneme features with spectrogram features, and a prosody predictor~\cite{Fastspeech2}, used to guide model training and enhance expressiveness. Recent advances in TTM are evident in models like MusicLM~\cite{agostinelli2023musiclm}, Noise2Music~\cite{huang2023noise2music}, MusicGen~\cite{copet2023simple-musicgen}, and MeLoDy~\cite{lam2023efficient-melody}. Similar to AudioLDM, the MusicLM model aligns music and language embeddings through contrastive pretraining modules, which enables text-free model optimization, alleviating the scarcity of music-text pairs. MusicLM also includes a semantic modeling stage based on w2v-BERT~\cite{chung2021w2v} to enhance the model performance. MusicGen uses a language modeling approach for music generation, enhanced with a mechanism for conditioning the model with melodic features for improved controllability. Meanwhile, MeLoDy, a diffusion model guided by language modeling, achieves significant computational reduction in music generation compared to MusicLM.

In this paper, we propose a unified framework for audio generation, which encompasses a breadth of topics including, but not limited to, speech, sound effects, and music generation.

\subsection{Diffusion Models}

\noindent
Diffusion models~\cite{DDPM, SGM} have demonstrated high sample quality in a variety of tasks including image generation~\cite{DiffusionBeatsGANs, DALLE2, Imagen}, image restoration~\cite{ISRIR}, speech generation~\cite{WaveGrad, DiffWave, leng2022binauralgrad}, and video generation~\cite{MakeAVideo, ImagenVideo}. In the realm of speech or audio synthesis, these models have been explored for both mel-spectrogram generation~\cite{popov2021gradtts, ResGrad} and waveform generation~\cite{BDDM, PriorGrad, InferGrad}. However, the iterative nature of generation in a high-dimensional data space often results in slow training and inference speeds. One solution involves the use of diffusion models in a more restricted latent space, a strategy exemplified in image generation~\cite{rombach2022high-stablediffusion}. 
This idea has been adopted in various audio generation works, including AudioLDM~\cite{liu2023audioldm}, Make-An-Audio~\cite{huang2023make-an-audio}, and TANGO~\cite{ghosal2023text-tango}. These works utilize latent diffusion models trained on a continuous latent space. On the other hand, there are also studies that explore diffusion in the discrete latent space. For instance, DiffSound~\cite{yang2022diffsound} employs a discrete autoencoder to mitigate redundancy~\cite{liu2023simple, liu2022learning} in the audio waveform and create a compressed representation of mel-spectrograms. DiffSound utilizes text-conditional discrete diffusion models to generate discrete tokens.

\section{AudioLDM 2} 
\label{sec: audiobox}

\subsection{Overview}
\label{sec: overview}
\begin{figure*}[t]
  \centering
  \includegraphics[width=0.9\textwidth]{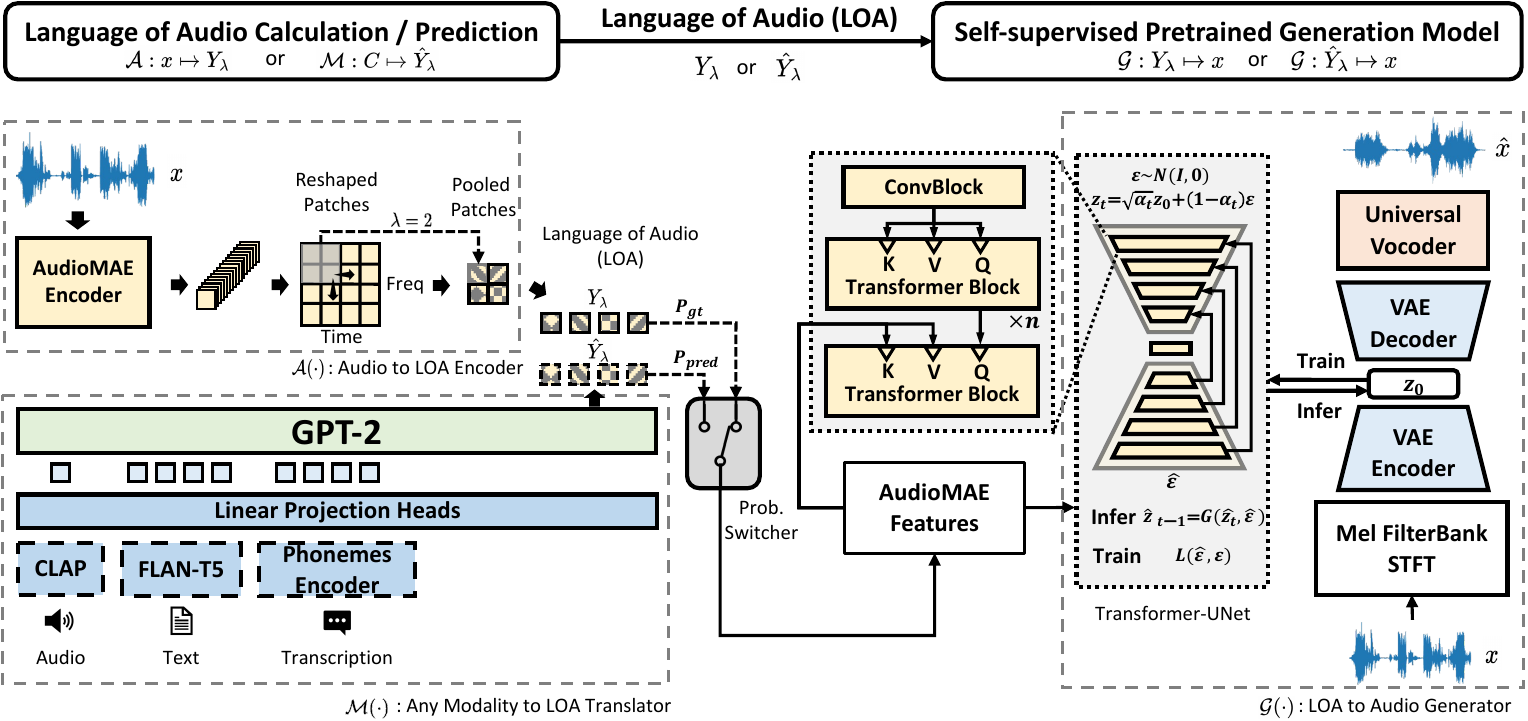}
  \caption{The overview of the \vModelName~architecture. The AudioMAE feature is a proxy that bridges the \textit{conditioning information to LOA translation} stage~(modelled by GPT-2) and the \textit{LOA to audio generation} stage~(modelled by the latent diffusion model). The probabilistic switcher controls the probability of the latent diffusion model using the ground truth AudioMAE~($P_{\text{gt}}$) and the GPT-2 generated AudioMAE feature~($P_{\text{pred}}$) as the condition. Both the AudioMAE and latent diffusion models are self-supervised pre-trained with audio data.}
  \label{fig:main-audiobox}
\end{figure*}

\noindent
Let $x\in \mathbb{R}^{L_{s}}$ represent an audio signal, where $L_{s}$ is the length of the audio samples in $x$. An audio generation process can be denoted as $\mathcal{H}: C\mapsto x$, where $C$ is the conditioning information and $\mathcal{H}$ is the conditional audio generation system. \


The direct generation of $x$ from $C$ is usually challenging~\cite{oord2016wavenet}. Motivated by regeneration learning~\cite{tan2023regeneration}, we propose to utilize an intermediate feature $Y$, as an abstraction of $x$, to bridge the gap between $C$ and $x$, as introduced in Section~\ref{sec:representation-learning-modules}. We call the feature $Y$ the language of audio~(LOA). The LOA feature is calculated by $Y=\mathcal{A}(x)$ 
in which $\mathcal{A}$ performs audio to LOA encoding with a self-supervised representation learning \textcolor{black}{module such as AudioMAE}~\cite{tan2023regeneration, xu2022masked}. 
\textcolor{black}{The LOA feature should be a representation that is potentially easier to model compared with $x$ and contain meaningful semantic information about $x$.}
As illustrated in Figure~\ref{fig:main-audiobox}, with the intermediate representation $Y$, the overall audio generation process can be denoted as 

\begin{equation}
    \mathcal{H}_{0}=\mathcal{G} \circ \mathcal{M} : C\mapsto \hat{Y} \mapsto x,
    \label{eq: overall-joint}
\end{equation}
where $\hat{Y}$ is the estimation of the ground truth LOA.
As denoted in~\eqref{eq: overall-joint}, the audio generation process of~\vModelName~includes the following two steps: 

\textit{(i) Conditioning information to LOA translation:}
The function $\mathcal{M}:C\mapsto\hat{Y}$ aims to produce the LOA $Y$ based on $C$, which could be the conditional information from \textcolor{black}{other} modalities, such as audio and text. As a potentially better representation of $C$\textcolor{black}{~in terms of audio generation}, the generated $\hat{Y}$ will be used in later stages as the conditioning information for audio generation. \textcolor{black}{We implement the function $\mathcal{M}$ with auto-regressive modelling, which is introduced in Section~\ref{sec: audio semantic language modeling}.}

\textit{(ii) LOA to audio generation:}
Followed by $\mathcal{M}$, function $\mathcal{G}$ accepts an LOA estimation $\hat{Y}$ as input condition and estimates the audio data $x$. 
During the training process, when the training data $x$ is available, the ground truth $Y$ will be also available using $\mathcal{A}(\cdot)$, allowing the optimization of $\mathcal{G}$ in a self-supervised manner. Specifically, instead of using the LOA estimation $\hat{Y}=\mathcal{M}(C)$, we condition the generation of $x$ based on the $Y=\mathcal{A}(x)$, which can be formulated as
\begin{equation}
    \mathcal{H}_{1}=\mathcal{G} \circ \mathcal{A} : x\mapsto Y \mapsto \hat{x}.
    \label{eq: ssl-pretrain}
\end{equation}

\noindent
We introduce the detail of $\mathcal{A}(\cdot)$ in Section~\ref{sec: audio-representation-learning}. Since the process $\mathcal{H}_{1}$ only involves $x$ as the training data, Equation~\eqref{eq: ssl-pretrain} means model $\mathcal{G}$ can be optimized in a self-supervised manner without any audio annotation. This self-supervised scheme can alleviate the scarcity of the audio data labels~\cite{liu2023audioldm} and provide a robust backbone for the overall generation system. \textcolor{black}{Note that the self-supervised learning here does not refer to the entire~\vModelName, for example, the function $\mathcal{M}$ still needs paired data to optimize. We implement the function $\mathcal{G}$ with the latent diffusion model, which is introduced in Section~\ref{sec: latent-diffusion-model}.}

The following sections provide a detailed introduction to \vModelName. 
In Section~\ref{sec: audio-representation-learning}, we discuss the audio representations employed in \vModelName, including the AudioMAE and VAE features. These features also serve as the generation targets for the two stages within \vModelName.
Section~\ref{sec: audio semantic language modeling} introduces the auto-regressive modeling of the AudioMAE feature with GPT-2. 
In Section \ref{sec: latent-diffusion-model}, we elucidate the process of generating audio waveforms via the latent diffusion model, which applies a VAE for feature compression and generates audio conditioned on the LOA. The LOA here can be based on either ground truth or GPT-2-generated data, which corresponds to self-supervised training and joint training with GPT-2~(Section~\ref{sec: joint-training}), respectively.


%





\subsection{Audio Representation Learning}
\label{sec: audio-representation-learning}


\textcolor{black}{Motivated by MusicLM~\cite{agostinelli2023musiclm} and AudioLM~\cite{borsos2023audiolm}, which perform semantic and acoustic modelling on two types of discrete representations~\cite{baevski2020wav2vec, zeghidour2021soundstream}, we adopt a similar two-stage modelling approach.  However, our work differs in that we work on the continuous semantic and acoustic representation, which can potentially provide richer information compared with discrete representations used by previous studies~\cite{zeghidour2021soundstream,chung2021w2v}. In our work, we adopt AudioMAE~\cite{xu2022masked} and variational autoencoder~(VAE)~\cite{kingma2013auto-vae} as the semantic and acoustic representation learning modules, respectively. Despite serving similar purposes, AudioMAE and VAE differ in architecture and objectives, yielding distinct representations. Further details on representation learning are provided below.}

\subsubsection{Semantic Representation Learning with the AudioMAE}
\label{sec:representation-learning-modules}

\noindent
To accurately represent diverse types of audio, encompassing speech, music, and sound effects, the LOA $Y$ should effectively capture both the semantic and the acoustic details of audio signals.
Therefore, we propose to use a self-supervised pretrained AudioMAE~\cite{xu2022masked} as the representation extraction module for function $\mathcal{A}$ for its generality and high accuracy on the downstream audio classification task~\cite{xu2022masked}. 

The audio masked autoencoder~(AudioMAE) is an audio self-supervised pre-training model, which learns representations from unlabeled audio data without relying on manually labeled annotations. An AudioMAE consists of an encoder and a decoder, both realized with an architecture similar to the vision transformer~(ViT)~\cite{dosovitskiy2020image}. During self-supervised pre-training, input patches to the encoder, which are usually mel spectrograms, are randomly masked and the decoder learns to reconstruct the masked patches~\cite{xu2022masked}. Compared with other audio self-supervised pretraining models, AudioMAE has the following two advantages: 

\textit{(i) The AudioMAE has been verified to work well in the general audio domain.} For example, an AudioMAE can be effectively pre-trained on AudioSet~\cite{gemmeke2017audio}, with state-of-the-art performance on the downstream audio classification tasks. In comparison, typical audio self-supervised models focus on a specific domain, such as the MERT~\cite{li2023mert} on music and the HuBERT~\cite{hsu2021hubert} on speech. 

\textit{(ii) AudioMAE features are potentially better for generative tasks than other discriminative pre-training methods.} Building upon the contrastive loss or next token prediction classification loss as the learning objective, previous systems such as wav2vec~\cite{schneider2019wav2vec} and BYOL-A~\cite{niizumi2021byol} utilize a discriminative approach during pre-training. In comparison, AudioMAE focuses on a generative process by learning the reconstruction of the masked patches. 

For an input audio signal $x$, AudioMAE first calculates the log mel spectrogram $X\in \mathbb{R}^{T\times F}$, where $T$ represents the time steps of the mel spectrogram, and $F$ denotes the mel bins. The mel spectrogram $X$ is then treated as an image and split into patches each of size $P\times P$, serving as the input for the AudioMAE encoder. The patch size $P$ is typically designed to be a common factor of $T$ and $F$. Patch splitting and embedding are performed using a convolutional neural network with a kernel size of $P$, a stride of $P$, and $D$ output channels. This yields an output shape of $T^{\prime}\times F^{\prime}\times D$, where $D$ is the AudioMAE embedding dimension, $T^{\prime}=T/P$, and $F^{\prime}=F/P$. The resulting output feature of the AudioMAE encoder, $E\in \mathbb{R}^{T^{\prime}\times F^{\prime}\times D}$, has the same shape as the input and is usually treated as the feature for downstream tasks after pretraining~\cite{xu2022masked}. 

\begin{figure}[t]
    \centering
    \includegraphics[width=1.0\linewidth]{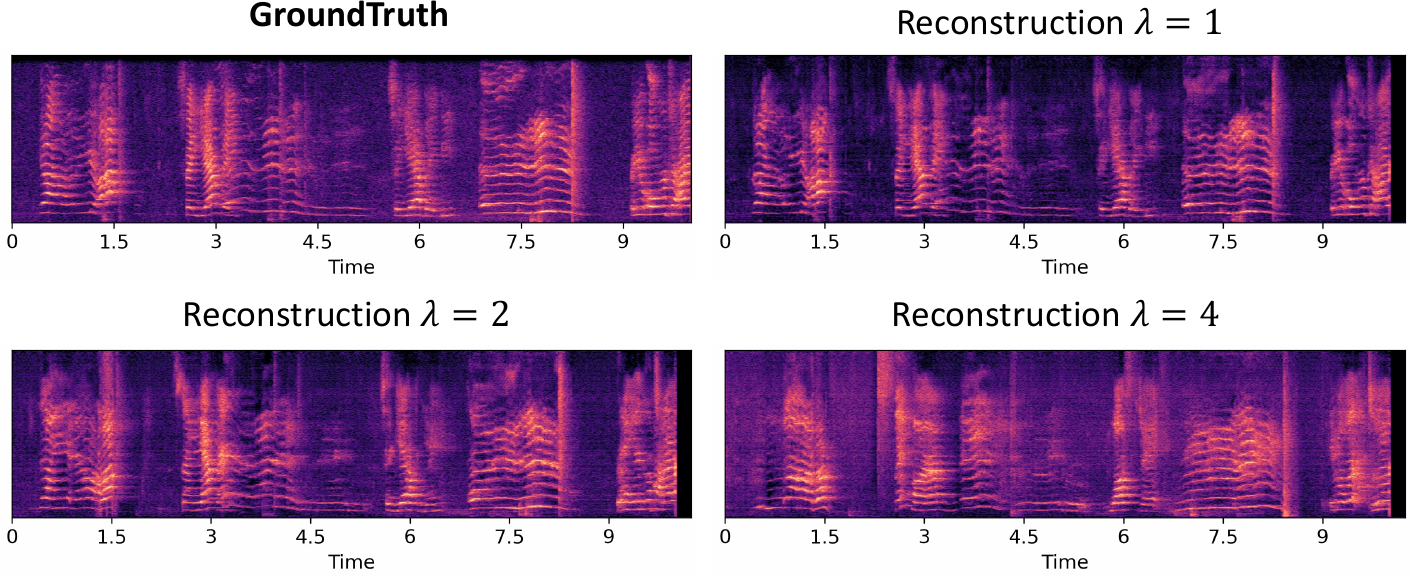}
    \caption{The influence of $\lambda$ on audio reconstruction from LOA $Y_{\lambda}$ \textcolor{black}{with the latent diffusion model}. The reconstruction closely resembles the ground truth when $\lambda=1$, suggesting that $Y_{\lambda=1}$ retains sufficient audio details. However, with $\lambda=2~\text{or}~4$, the reconstruction diverges slightly from the original audio, indicating that while the post-processed AudioMAE feature may not include all details, it nonetheless accurately preserves semantic content.}
    \label{fig: effect-of-lambda_compressed}
\end{figure}

\subsubsection{AudioMAE Feature Post Processing}

As shown in Figure~\ref{fig:main-audiobox}, once the AudioMAE features $E$ are computed, we introduce an additional pooling step to aggregate $E$ into $Y_{\lambda}$, where $\lambda\in I^{+}$ represents a hyper-parameter used in the post-processing pooling step. This pooling step aims to reduce the sequence length, facilitating easier estimation in the function $\mathcal{M}$.
Specifically, we perform a two-dimensional average-max pooling~\cite{liu2023simple} on the first two dimensions of $E\in \mathbb{R}^{T^{\prime}\times F^{\prime}\times D}$, in which the pooling kernel size and stride have the same value $\lambda\in I^{+}$. 
The two-dimensional pooling operation can help to preserve the time-frequency relationship in the output. 
The final output after pooling, $Y_{\lambda}$, is reshaped into a embedding sequence with shape $L_{\lambda}\times D$, in which $L_{\lambda}=T^{\prime}F^{\prime}/\lambda^{2} $. To facilitate implementation, $\lambda$ is chosen so that $L_{\lambda}$ is always a positive integer. We demonstrate the effect of different choices of $\lambda$ in Figure~\ref{fig: effect-of-lambda_compressed}. In the remaining sections of this paper, if $\lambda$ is not specified, we'll refer to $Y_{\lambda}$ simply as $Y$.

\subsubsection{Acoustic Representation Learning with VAE}  

We use a VAE for feature compression and for learning an audio representation $z$, which has a significantly smaller dimension than $x$~\cite{liu2023audioldm}. The VAE we used in this work is a convolutional architecture that consists of encoders with down-sampling and decoders with up-sampling following the architecture described in~\cite{liu2023audioldm}. The forward pass of the VAE can be formulated as $\mathcal{V}: X\mapsto z\mapsto \hat{X}$, where $X$ is the mel-spectrogram of $x$ and $\hat{X}$ is the reconstruction of $x$. The reconstruction $\hat{X}$ can be converted to the audio waveform $\hat{x}$ using a pre-trained HiFiGAN vocoder~\cite{kong2020hifi}. Following AudioLDM~\cite{liu2023audioldm}, we calculate a reconstruction loss and a discriminative loss based on $X$ and $\hat{X}$ to optimize the parameters of the VAE. We also calculate the KL divergence between $z$ and a standard Gaussian~($\mu=0,~\sigma^{2}=1$) as a loss function to limit the variance of the VAE latent space.

\subsubsection{Comparison between AudioMAE and VAE} 

\begin{figure}
    \centering
    \includegraphics[width=1.0\linewidth]{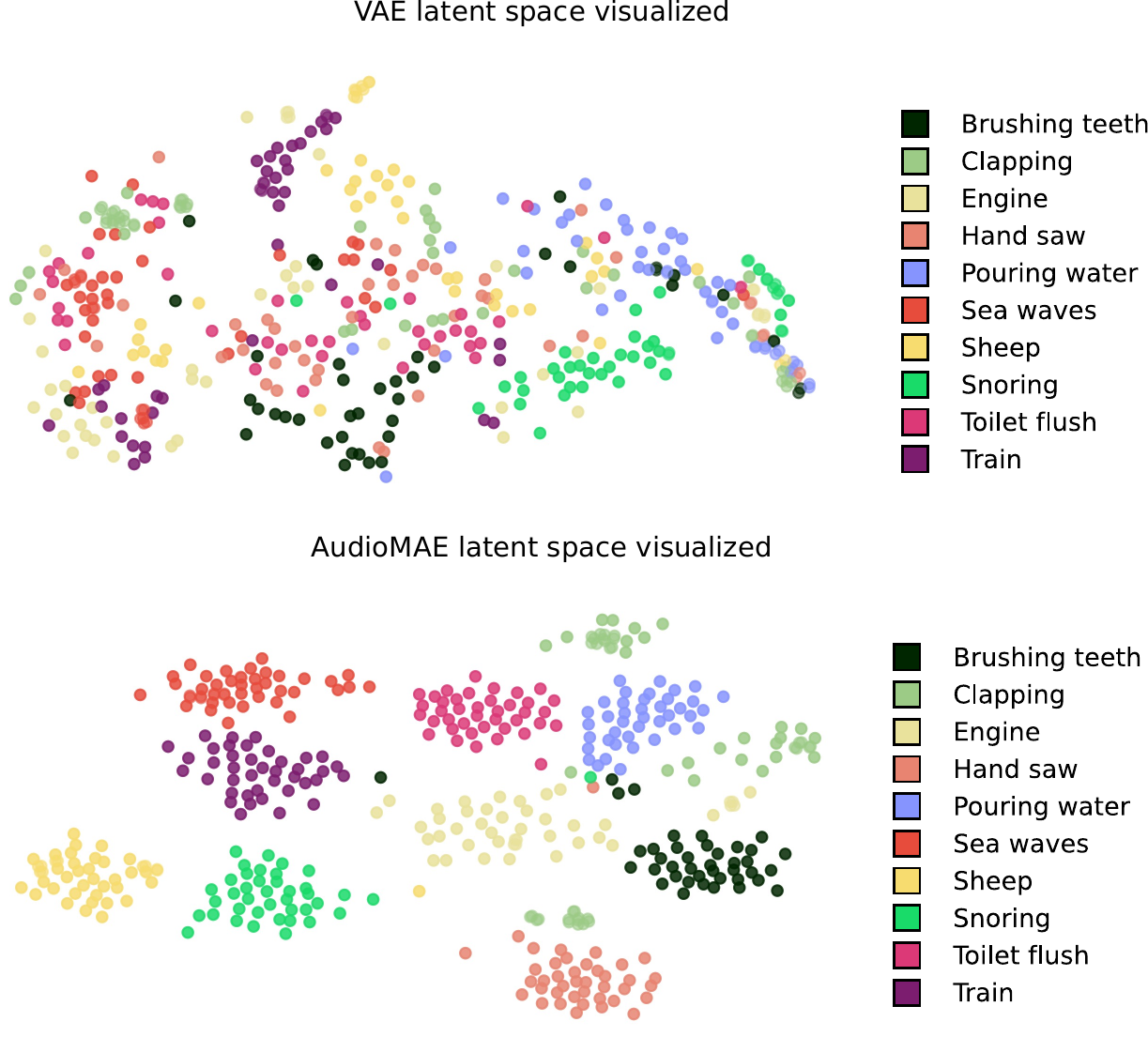}
    \caption{Visualization of the latent space based on tSNE and ten randomly selected classes in the ESC50~\cite{piczak2015esc} dataset. Each point in the figure represents an audio clip. The AudioMAE feature space tends to group similar audio clips together, indicating more semantic structure than in the VAE feature.}
    \label{fig:tsne-comparison-visualization}
\end{figure}

Since both AudioMAE and VAE are based on autoencoders for representation learning, one might wonder why we use a VAE for representation learning instead of directly modeling the AudioMAE latent space.
Part of the reason is that AudioMAE does not primarily focus on reconstruction quality, and its latent space compression ratio is not as high as that of the VAE. On the other hand, the VAE exhibits good reconstruction ability and a higher compression level than AudioMAE, making VAE more suitable for mel-spectrogram compression.
Furthermore, as shown in Figure~\ref{fig:tsne-comparison-visualization}, we visualize the latent representation of AudioMAE and VAE on the ESC-50~\cite{piczak2015esc} dataset using tSNE~\cite{van2008visualizing-tsne}. The visualization demonstrates that the latent representation of AudioMAE can group similar audio at a closer region in the latent space. In contrast, the representation of VAE exhibits more overlap between different audio classes. This indicates that the representations for the AudioMAE and VAE are distinct. AudioMAE contains more information on the semantic side, while VAE representation is less semantically structured. \textcolor{black}{Therefore according to the definition of LOA in Section~\ref{sec: overview}, AudioMAE is more suitable than VAE on calculating the LOA.}

\subsection{Conditioning Information to LOA Translation with GPT-2}
\label{sec: audio semantic language modeling}

\noindent
This subsection introduces the design of the function $\mathcal{M}$. As introduced in Section~\ref{sec: overview}, the input to the model $\mathcal{G}: Y \mapsto x$ can be calculated using the AudioMAE. However, during inference, when we perform audio generation with the condition $C$, the ground truth LOA $Y=\mathcal{A}(x)$ is unavailable. Therefore, we need another model that can generate $\hat{Y}$ given $C$, denoted by $\mathcal{M}_{\theta}: C\rightarrow \hat{Y}$, where $\theta$ represents trainable parameters. 


Specifically, we treat the generation of $Y$ as a language modelling task and choose the GPT-2~(Generative Pre-trained Transformer 2)~\cite{radford2019language} model as the backbone. GPT-2 is based on a transformer architecture and was originally trained on $8$ million documents for a total of $40$~GB of text using an unsupervised learning approach~\cite{radford2019language}. 
GPT-2 has been used in a variety of natural language processing tasks, such as text completion, question answering, and language translation~\cite{qu2020text,klein2019learning}. 
Initialized with pre-trained weights, we finetune the GPT-2 model based on teacher forcing~\cite{lamb2016professor}, so that during model training, $\hat{y}_{l}$ will be generated based on both the condition $C$ and the ground truth sequence $y_{1},...,y_{l-1}$, where $y_l$ is the $l-$th vector in LOA sequence $Y$. Specifically, the GPT-2 model $\mathcal{M}_{\theta}$ is trained to maximize the likelihood of a sequence $Pr(y_1,y_2,...,y_{L}|C)$, which can be interpreted into the following optimization objective:
\begin{equation}
    \label{eq: autoregressive-objective}
    \text{argmax}_{\theta}~\mathbb{E}_{C}~[Pr(y_1|C; \theta)\prod_{l=2}^{L}Pr(y_{l}|y_1,...,y_{l-1}, C; \theta)],
\end{equation}
where $\mathbb{E}_{C}$ represents the expectation operator with respect to the variable $C$.
We calculate the mean squared error loss~\cite{oord2016wavenet} between $y_{l}$ and $\hat{y}_{l}=\mathcal{M}_{\theta}(y_1,...,y_{l-1}, C)$ to optimize Equation~\eqref{eq: autoregressive-objective}. We directly optimize the regression of continuous vectors $y_{l}$, without discretizing the AudioMAE feature space and estimating the token index. 
The condition $C$ in Equation~\eqref{eq: autoregressive-objective} can encompass a flexible range of data representations, including audio representations, text embeddings, phoneme embeddings, or visual clues. We adopt the mixture of experts~\cite{masoudnia2014mixture} approach and use multiple encoders as feature extractors to calculate $C$. Given $K$ systems as the feature extraction modules, the shape of the output from the $k$-th system $C_{k}, k\in\{1,...,K\}$ is $L_{k}\times D_k$, in which $L_{k}$ is the sequence length of the $k$-th system and $D_k$ is the dimension of the feature. We apply a linear transformation layer after the output of each feature extraction module to unify the embedding dimension to $D_0$ for an easier process of the GPT-2 model. For modules that extract global features from the input without sequential information, such as CLAP~\cite{wu2023large-clap} or ImageBind~\cite{girdhar2023imagebind}, we have $L_{k}=1$. The final condition $C=[C_{1},...C_{K}]$ is a concatenation of $C_{k}$ along the sequence length dimension. The final condition $C$ has a shape of $L\times D_0$, where $L=\sum_{k=1}^{K}L_{k}$. We introduce several condition modules we used in this paper as follows.  

\noindent
\textbf{CLAP} or contrastive language and audio pretraining~\cite{wu2023large-clap}, is a system that learns a joint audio-text embedding space, in which paired audio and language data have a closer distance in the latent space. CLAP has been successfully applied as a conditioning module to audio generation such as AudioLDM~\cite{liu2023audioldm}. In this study, we employ a pre-trained CLAP\footnote{\url{https://github.com/LAION-AI/CLAP}} text encoder as the default conditioning module for extracting text embeddings as conditions. However, in scenarios where text captions~(e.g., ``A man is speaking happily with background static noise'') are unavailable, such as for text-to-speech tasks, we use the CLAP audio encoder as the conditioning module instead of using CLAP text encoder, in the same way as~\cite{liu2023audioldm}.


\noindent
\textbf{FLAN-T5.} The CLAP model, as a module that calculates global-level conditions, has been found to have issues in capturing the temporal information in the text data~\cite{wu2023audio}. To allow for this, we use another pretrained text encoder to capture the semantic information of the textual input, which might contain useful details such as temporal orders. Specifically, we utilize FLAN-T5~\cite{chung2022scaling-flan-t5}, which is an enhanced version of the text-to-text transfer transformer~(T5) model~\cite{raffel2020exploring} based on the finetuning on a mixture of tasks\footnote{\url{https://huggingface.co/google/flan-t5-large}}. 

\noindent
\textbf{Phoneme Encoder} is a widely adopted module in text-to-speech research for extracting helpful information regarding phonemes~\cite{Fastspeech2, tan2022naturalspeech}, which are the smallest units of sound in a language that can distinguish one word from another~\cite{tan2023neural}. In this work, we follow the structure introduced in NaturalSpeech~\cite{tan2022naturalspeech} to build a phoneme encoder, in the form of a stack of transformer encoder layers. We preprocess the textual input into phonemes using the open-source tool Espeak phonemizers~\footnote{\url{https://github.com/espeak-ng/espeak-ng}} and append a stop token after each phoneme sequence to mark the end of the sequence for the transformer model. 




Except for the phoneme encoder, which does not have a readily available pre-trained weights, the parameters of all other pre-trained feature extraction models are kept frozen during the experiment.

\subsection{LOA to Audio Generation with Latent Diffusion Model} 
\label{sec: latent-diffusion-model}

\noindent
We model the process $\mathcal{G}: Y \mapsto x$ with a latent diffusion model~(LDM)~\cite{rombach2022high-stablediffusion}, which is a variant of the denoising diffusion probabilistic models~(DDPM)~\cite{DDPM}. In contrast to DDPM, which directly models the training data, the LDM learns the reverse diffusion process in a variational autoencoder~(VAE)-based compressed latent space~\cite{kingma2013auto}, which can reduce the computational cost. Similar ideas have been adapted to audio generation, such as AudioLDM~\cite{liu2023audioldm}.

\subsubsection{Latent Diffusion Model}
\noindent
We follow the formulation in ~\cite{DDPM} to implement the LDM. Given a VAE representation $z$, the forward transition is defined as a $T$ steps Markov process in a way that does not include trainable parameters. Given the data $z_{t-1}$ at diffusion step $t-1$, the data distribution of $z_{t}$ at step $t\in{2,...,T}$ can be formulated as

\begin{equation}
    \label{eq: ddpm-t-1-to-t-q-distribution}
    q(z_{t}|z_{t-1})=\sqrt{1-\beta_{t}}z_{t-1}+\sqrt{\beta_{t}}\epsilon_{t}, 
\end{equation}

\noindent
in which the noise schedule hyper-parameter $\beta_{t}\in[0,1]$ determines how quickly the noise is blended into the data. By recursive substitution of $q(z_{t}|z_{t-1})$ in Equation~\eqref{eq: ddpm-t-1-to-t-q-distribution}~\cite{DDPM},  we can derive the distribution of $z_{t}$ given $z_0$ as

\begin{equation}
    \label{eq: close-form-q(t)}
    q(z_{t}|z_{0})=\sqrt{\alpha_{t}}z_0+\sqrt{1-\alpha_{t}}\epsilon_t, 
\end{equation}

\noindent
where $\alpha_{t}=\prod_{t=1}^{t}1-\beta_{t}$ and $\epsilon_{t}\sim N(0,I)$ . At the final step $t=T$, the distribution of $z_{t}$ will be close to a standard Gaussian distribution~\cite{DDPM}. 

The LDM learns a backward transition from the prior distribution $N(0,I)$ to the data distribution $z$. The reverse process models the conditional distribution $Pr(z_{0...T}| Y; \phi)=Pr(z_0| z_1, Y; \phi)\prod_{t=2}^{T}Pr(z_{t-1}|z_{t}, Y; \phi) \cdot Pr(z_T)$, in which $Y$ is the LOA as the condition signal and the $\phi$ denotes the parameter of the model for learning the reverse diffusion. If we marginalize $z_{1...T}$ we can derive the lower bound of $\text{log}[Pr(z_0| Y; \phi)]$ based on the evidence lower bound~(ELBO) and Bayes' rule~\cite{DDPM}:

\begin{align}
    \label{eq: ELBO-lower-bound}
    \text{log}[Pr(z_0| Y; \phi)] &\geq \text{log}[Pr(z_0|z_1, Y; \phi)] \nonumber - \\
    &\quad \hspace{-4.5em} \sum_{t=2}^{T}KL[Pr(z_{t-1}|z_{t}, Y; \phi) || q(z_{t-1}|z_t, z_0)],
\end{align}

\noindent
where $KL(\cdot)$ is the function for calculating KL divergence, and $q(z_{t-1}|z_t, z_0))$ is the target conditional diffusion distribution that has a closed-form solution given $z_0$ and $z_t$~\cite{DDPM}.  Following~\cite{DDPM}, we can derive the loss function that maximizes the lower bound of Equation~\eqref{eq: ELBO-lower-bound} as:

\begin{equation}
    \label{eq: latent-diffusion-model-loss}
    \text{argmin}_{\phi}[\mathbb{E}_{z_0,~Y,~t\sim \{1,...,T\}}|| \mathcal{G}(\sqrt{\alpha_{t}}z_0+\sqrt{1-\alpha_{t}}\epsilon_t, t, Y; \phi)-\epsilon_t ||].
\end{equation}
As shown in Figure~\ref{fig:main-audiobox}, we \textcolor{black}{utilize} a Transformer-UNet~(T-UNet) architecture as the function $\mathcal{G}$ in Equation~\eqref{eq: latent-diffusion-model-loss}, which \textcolor{black}{is similar} to the UNet used in AudioLDM~\cite{liu2023audioldm} \textcolor{black}{but with more transformer layers}. 
The T-UNet architecture consists of a series of encoders with downsampling and a series of decoders with upsampling, and there are skip connections between encoders and decoders at the same scale. To enhance the modelling capacity of the T-UNet, we insert multiple transformer blocks after the convolution operation in each encoder and decoder block. Specifically, we have $n_{\text{trans}}+1$ transformer blocks, in which the first $n_{\text{trans}}$ transformer blocks are a stack of self-attention layers~\cite{vaswani2017attention-is-all-you-need} and feed-forward networks. To incorporate the condition information $Y$ from the ground truth LOA or $\hat{Y}$ from $\mathcal{M}(\cdot)$~(Section~\ref{sec: audio semantic language modeling}), as shown in Figure~\ref{fig:main-audiobox}, the last transformer block changes the self-attention layer to cross-attention, which accepts the LOA as key and value and fuses with the feature from the previous transformer block as the query. Except for text-to-speech generation, we add an extra cross-attention layer in the transformer block to accept the text embedding from FLAN-T5~\cite{chung2022scaling-flan-t5} as an extra condition to enhance the audio-text relationship learning. 


\subsubsection{Classifier-free Guidance}
\label{CFG}

\begin{figure}[htbp]
    \centering
    \includegraphics[width=1.0\linewidth]{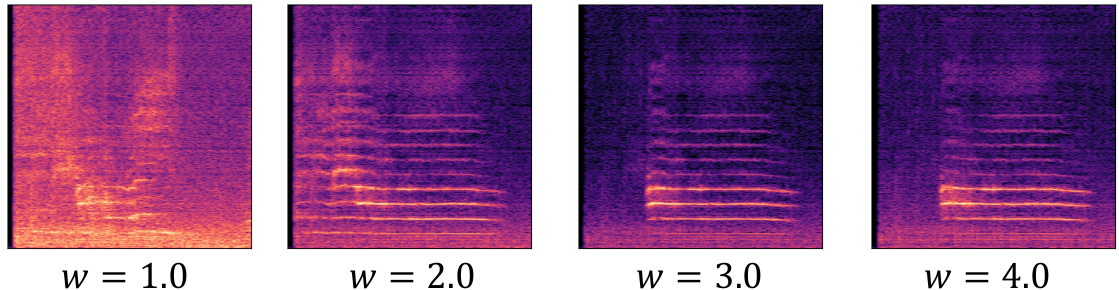}
    \caption{The samples generated with different classifier-free guidance scales. The text prompt is \textit{``A cat is meowing"}.}
    \label{fig:my_label}
\end{figure}

For diffusion models, controllable generation can be achieved by introducing guidance at each sampling step. Classifier-free guidance~\cite{CFG,Glide} (CFG) has been the state-of-the-art technique for guiding diffusion models.
During training, we randomly discard our condition $Y$ in Equation~\eqref{eq: latent-diffusion-model-loss} with a fixed probability (e.g., $10\%$) to train both the conditional LDMs $\mathcal{G}(z_{t}, t, Y; \phi)$ and the unconditional LDMs $\mathcal{G}(z_{t}, t, \phi)$. For generation, we use LOA $\hat{Y}$ or $Y$ as the condition and perform sampling with a modified noise estimation $\mathcal{G^{\prime}}(z_{t}, t, Y; \phi)$:
\begin{align}
\mathcal{G^{\prime}}(z_{t}, t, Y; \phi) = w\mathcal{G}(z_{t}, t; \phi)+(1-w)\mathcal{G}(z_{t}, t, Y; \phi),
\end{align}
where $w$ determines the guidance scale.


\subsubsection{Joint Finetuning}

\label{sec: joint-training}
We perform joint finetuning with the GPT-2 and latent diffusion models based on Equation~\eqref{eq: overall-joint},~\eqref{eq: latent-diffusion-model-loss}, and \eqref{eq: autoregressive-objective}. As demonstrated by Table~\ref{tab: ablation}, we found that joint finetuning significantly enhances the overall performance of the~\vModelName~system. As depicted in Figure~\ref{fig:main-audiobox}, the \textit{probabilistic switcher} controls the source of the conditioning signal during the joint training process. During training, the switcher dynamically chooses between ground truth AudioMAE features and GPT-generated AudioMAE features, with probabilities set to $P_{\text{gt}}$ and $P_{\text{pred}}$, respectively.

\section{Experiment Setup}

\subsection{Dataset} 
\noindent
\label{sec: dataset}
The datasets used in this work include AudioSet~(AS)~\cite{gemmeke2017audio}, WavCaps~\cite{mei2023wavcaps}, AudioCaps~(AC)~\cite{kim2019audiocaps}, VGGSound~(VS)~\cite{chen2020vggsound}, Free Music Archive~(FMA)~\cite{defferrard2016fma}, Million Song Dataset~(MSD)~\cite{bertin2011million}, LJSpeech~(LJS)~\cite{ljspeech17}, and GigaSpeech~(GGS)~\cite{chen2021gigaspeech}. AudioSet is the largest audio classification dataset at the time of writing, with around two million ten-seconds of audio and $527$ different classes. 
WavCaps is a dataset with ChatGPT-assisted weakly-labeled audio captions. WavCaps contains 403,050 audio clips with an average duration of 68 seconds. AudioCaps is a subset of AudioSet with handcrafted captions, containing about 46,000 ten-second audio clips. VGGSound is a large-scale single-label audio-visual dataset, which contains over 200,000 videos. We only utilize the audio data and the labels in the VGGSound. FMA is a large music dataset without captions, containing 106,574 music tracks from 16,341 artists and 14,854 albums. 
For the Million Song Dataset, we only utilize the labelled subset proposed in~\cite{toward2022doh-ecals}, which contains around 510,000 music tracks with metadata such as tags, titles, and artist names. LJSpeech is a single-speaker speech dataset with 13,100 short audio clips and detailed transcriptions. 
GigaSpeech is a multi-speaker large-scale English speech recognition corpus with around 10,000 hours of audio labeled with transcriptions. The test and development set of GigaSpeech are not included during training. All the audio data used in this work are resampled to $16$ kHz for easier comparison with previous works~\cite{liu2023audioldm, huang2023make-an-audio}. We use only the audio data with paired text labels to train the GPT-2 model by optimizing Equation~\eqref{eq: autoregressive-objective}. We train the latent diffusion model with all the audio data regardless of annotation by optimizing the objective in Equation~\eqref{eq: ELBO-lower-bound} in a self-supervised manner.

\subsection{Evaluation Metrics} 
\noindent
We mainly focus on the text-to-audio generation task to evaluate the effectiveness of \vModelName. We follow the evaluation protocol of AudioGen~\cite{kreuk2022audiogen}, which calculates both objective metrics such as Frechet Audio Distance~(FAD), Kullback-Leibler Divergence~(KL), and subjective metrics including Overall Impression~(OVL) and Audio and Text Relation~(REL). 
\textcolor{black}{We also include an additional metric CLAP score~\cite{huang2023make-an-audio} to measure the correspondancy between the generated audio and text prompt.}
FAD is a reference-free audio quality measure that is calculated based on the distribution distance between the feature of the target and generated audios, extracted from the VGGish~\cite{vggish_hershey2017cnn} model. 
KL divergence measures the similarity between the generated and target audio with the label calculated by the audio tagging model, Patch-out Transformer~\cite{koutini2021efficient}, in the same way as AudioGen~\cite{kreuk2022audiogen}. 
\textcolor{black}{CLAP score measures the similarity between audio and text based on a pair of pretrained audio and text encoders~\cite{wu2023large-clap}, given by}

\begin{equation}
    \textcolor{black}{\text{CLAPScore}(x, r) = \frac{\vec{e}_{x} \cdot \vec{e}_{r}}{\text{max}(\| \vec{e}_{x} \| \| \vec{e}_{r} \|, \epsilon)},}
\end{equation}

\noindent
\textcolor{black}{where $x$ and $r$ denote audio and text data, respectively, $\epsilon$ is a small value that can avoid zero division, $\vec{e}_{a}$ is the output of the CLAP audio encoder and $\vec{e}_{r}$ is the output of CLAP text encoder. The value range of the CLAP score is between $-1$ and $1$ and a larger value indicates a stronger correlation between audio and text information.}

We use a similar evaluation protocol for text-to-music generation. For the text-to-speech task, we utilize the commonly used mean opinion score~(MOS) for evaluation~\cite{tan2023neural}. 

\subsection{Subjective Evaluation} 
\noindent
We use Amazon Mechanical Turk\footnote{\url{https://requester.mturk.com/}}, a crowd-sourced platform, to evaluate subjective metrics including OVL, REL, and MOS. The instructions on how to perform evaluation are clearly illustrated for the raters with examples. \textcolor{black}{Specifically, for OVL, raters were asked \textit{How would you rate the overall quality of this music? Consider its resemblance to real-world audio and its naturalness,} with a five-point scale ranging from \textit{5-Excellent quality} to \textit{1-Bad quality}. Similarly, for REL, the question posed was, \textit{How would you rate the relevance of music to the text description?} with a similar five-point scale for responses. In evaluating MOS, the question was, \textit{How natural does this recording sound? Take into account emotion, prosody, and other human-like details,} with options ranging from \textit{completely unnatural speech} to \textit{perfectly natural speech}.} To ensure the credibility of the evaluation result, we set requirements for the crowd-source worker with a minimum average approval rate of $60\%$ and with at least $50$ approvals in the record. 
Each audio clip is evaluated by at least $10$ different raters. All three subjective metrics have a Likert scale~\cite{likert1932technique} between one and five, where a larger number indicates better performance. Study raters received payment at or above the US minimum wage. We average the scores among all raters and samples as the final score for a system. 

\subsection{Model Architecture Details} 
\noindent

We perform the experiment with two sizes of the latent diffusion model, \vModelName~and \vModelName\textit{-Large}, with transformer layer numbers $n_{\text{trans}}=2$ and $n_{\text{trans}}=6$~(Section~\ref{sec: latent-diffusion-model}), respectively. We use a pre-trained AudioMAE\footnote{\url{https://github.com/facebookresearch/AudioMAE}} with a patch size of $16\times 16$ and no overlapping, resulting in a $768$-dimension feature sequence with length $512$ for every ten seconds of mel spectrogram. In a similar way to the idea introduced in~\cite{chen2023speech}, on calculating the LOA $Y$, we gather the output of the last $16$ transformer layers from the AudioMAE encoder and perform averaging as the final $Y$. 
\textcolor{black}{We perform self-supervised pre-training on both \vModelName~and \vModelName\textit{-Large} with the audio data mentioned in Section~\ref{sec: dataset}.}
The GPT-2 model we employ has an embedding dimension of $768$ with $12$ layers of transformers. For joint fine-tuning, we set the probability of using ground truth LOA $Y$ and LOA estimation $\hat{Y}$ as $P_{\text{gt}}=0.25$, and $P_{\text{pred}}=0.75$, respectively.

\begin{table}[htbp]
\textcolor{black}{
    \caption{The setup of the primary experiments we performed. FULL represents a combination of five different datasets, including AC, AS, WC, VS, and MSD. The model with $\dag$ use CLAP text encoder and FLAN-T5 to calculate conditions while the model with $\ddagger$ uses CLAP audio encoder and the phoneme encoder as the conditional modules.}
    \label{tab: model-setups}
    \begin{tabular}{lcccc}
    \toprule
    Model                     & $\lambda$ & Param & Dataset      & Task    \\
    \midrule
    AudioLDM 2-AC$^\dag$             & $8$                      & $346$M    & AC                & TTA     \\
    AudioLDM 2-MSD$^\dag$            & $8$                      & $346$M    & MSD               & TTM     \\
    AudioLDM 2-Full$^\dag$           & $8$                      & $346$M    & FULL   & TTA/TTM \\
    AudioLDM 2-AC-Large$^\dag$       & $8$                      & $712$M    & AC                & TTA     \\
    AudioLDM 2-Full-Large$^\dag$     & $8$                      & $712$M    & FULL & TTA/TTM \\
    AudioLDM 2-LJS$\ddagger$            & $1$                      & $346$M    & LJS               & TTS     \\
    AudioLDM 2-LJS-Pretrained$\ddagger$ & $1$                      & $346$M    & LJS+GGS           & TTS  \\
    \bottomrule
    \end{tabular}
}
\end{table}

\begin{figure}
    \centering
    \includegraphics[width=1.0\linewidth]{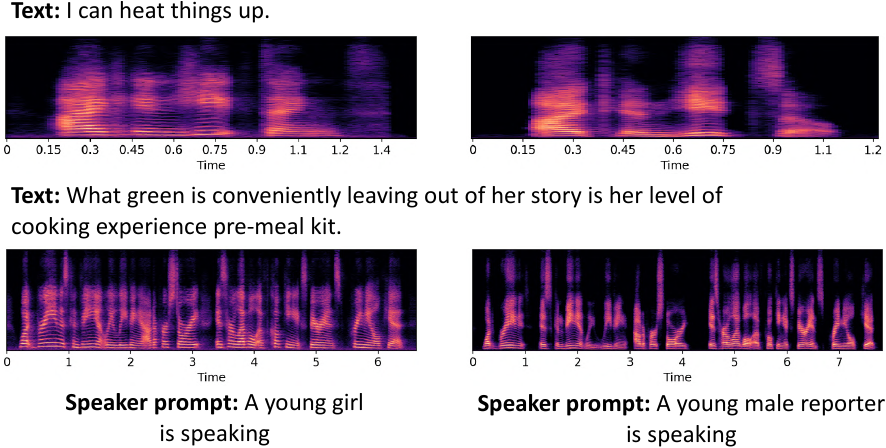}
    \caption{Examples of speaker-prompted text-to-speech generation. We use speaker prompts to describe the characteristics of the speaker and provide the model with the text transcription. }
    \label{fig: speaker-control}
\end{figure}

\textcolor{black}{Table~\ref{tab: model-setups} summerize the experiments we performed in this paper.} For the generation of audio and music, we combine the text embeddings from the CLAP text encoder and FLAN-T5 as conditioning and designate $Y_{\lambda=8}$ as the target sequence for GPT. The conditioning modules for speech generation are configured differently, primarily due to the need to better preserve the fine-grained phoneme information in speech signals through a smaller $\lambda$ value. Thus, for speech generation, we \textcolor{black}{concatenate the output of CLAP audio encoder and the phoneme encoder as the input sequence of the GPT-2 model}, and designate $Y_{\lambda=1}$ as the target sequence to retain more details. For the speech data, since there are no available audio captions~(different from transcriptions), we adopt a similar approach as AudioLDM~\cite{liu2023audioldm} to utilize the CLAP audio encoder to compute the embedding as a condition during model training, and employ the CLAP text encoder during inference. This method also facilitates prompt-based speaker control, as demonstrated in Figure~\ref{fig: speaker-control}.

\subsection{Training and Inference Setup}
\noindent

The latent diffusion model and the GPT-2 model are initially trained separately. We randomly choose $\lambda\in\{1,2,4,8\}$ during pre-training of the latent diffusion model to enhance the model robustness under conditions $Y_{\lambda}$ with different $\lambda$. \textcolor{black}{$Y_{\lambda}$ is only used as key and value in the T-UNet cross-attention layers therefore $Y_{\lambda}$ can have varying length.} \textcolor{black}{We train the latent diffusion model based on $10$ seconds of random segment from the training set. For easier modeling of the T-UNet, we zero-pad the $10$ seconds of audio segment into $10.24$ seconds during model training.}
We train the latent diffusion model and finetune the GPT-2 model on eight NVIDIA A100 80GB GPUs. \textcolor{black}{We follow the settings described in AudioLDM~\cite{liu2023audioldm} and} change the default classifier-free guidance scale during the Denoising Diffusion Implicit Models~(DDIM)~\cite{song2020denoising-ddim} sampling to $3.5$.
For both GPT-2 finetuning and the latent diffusion model, we utilize the AdamW~\cite{loshchilov2017decoupled} optimizer with a learning rate of $10^{-4}$ and $10000$ steps of linear warming up without decay.

\begin{table*}[tbp]
\centering
\small
\caption{Comparison of model performances on the AudioCaps evaluation set. \textcolor{black}{GT-AudioMAE denote directly applying the ground truth ``Language of Audio'' $Y$, to the function $\mathcal{G}$ for audio generation, as detailed in Section~\ref{sec: overview}}. \vModelName~significantly surpasses previous methods in both subjective and objective assessments. \textcolor{black}{All models are trained using the AudioCaps training subset. Models marked with $\ast$ are exclusively trained on this subset, while those with $\#$ are fine-tuned on it. }}
\begin{tabular}{lccccccc}
\toprule
Model              & Duration (h) & Param & FAD$\downarrow$           & KL$\downarrow$            & CLAP$\uparrow$     & \multicolumn{1}{c}{OVL}$\uparrow$ & \multicolumn{1}{c}{REL}$\uparrow$ \\
\midrule
GroundTruth        & -            & -          & -             & -             & $0.251$          & $4.04$ & $4.08$ \\
\textcolor{black}{GT-AudioMAE}        & -            & -          & $1.84$             & $0.19$             & $0.239$          & $3.87$ & $4.02$ \\
AudioGen-Large     & $6824$         & $1$~B         & $1.82$          & $1.69$          & -              & \multicolumn{1}{c}{-}    & \multicolumn{1}{c}{-}    \\
Make-an-Audio      & $3000$         & $453$~M       & $2.66$          & $1.61$          & -              & \multicolumn{1}{c}{-}    & \multicolumn{1}{c}{-}    \\
AudioLDM-Large$^\#$      & $9031$      & $739$~M       & $1.96$          & $1.59$          & -              & \multicolumn{1}{c}{-}    & \multicolumn{1}{c}{-}    \\
AudioLDM-M         & $9031$      & $416$~M       & $4.53$          & $1.99$          & $0.141$          & $3.61$  & $3.55$ \\
Make-an-Audio 2   & 3700         & $937$~M       & $2.05$          & $1.27$          & $0.173$          & $3.68$  & $3.62$ \\
TANGO$^\ast$              & $145$          & $866$~M       & $1.73$          & $1.27$          & $0.176$          & $3.75$ & $3.72$ \\
\midrule
\vModelName\textit{-AC}$^\ast$           & $145$          & $346$~M       & $1.67$          & $1.01$          & $\mathbf{0.249}$          & $3.88$ & $\mathbf{3.90}$ \\
\vModelName\textit{-Full}    & $29510$          & $346$~M  & $1.78$                 & $1.60$                 & $0.191$                &  $3.83$                       &           $3.77$  \\
\vModelName\textit{-AC}-Large$^\ast$     & $145$          & $712$~M      &   $\mathbf{1.42}$              &        $\mathbf{0.98}$       &         $0.243$       & $\mathbf{3.89}$ & $3.87$ \\
\vModelName\textit{-Full-Large}    & $29510$          & $712$~M       & \multicolumn{1}{c}{$1.86$} & \multicolumn{1}{c}{$1.64$} & \multicolumn{1}{c}{$0.182$} &   $3.79$                      &      $3.80$        \\
\midrule
\end{tabular}
\label{tab: main-audiocaps_sota-comparison}
\end{table*}

\section{Result}

\noindent
We evaluated our proposed system on three primary audio generation tasks: text-to-audio, text-to-music, and text-to-speech. The three basic systems were trained on three different datasets: AudioCaps~(general audio), MSD~(music), and LJSpeech~(speech), and are denoted as \vModelName\textit{-AC}, \vModelName\textit{-MSD}, and \vModelName\textit{-LJS}, respectively. The model \vModelName\textit{-Full} represents a version capable of performing both audio and music generation simultaneously, with training data scaled up to $29510$ hours\textcolor{black}{, including all available data mentioned in Section~\ref{sec: dataset}}. In contrast with AudioLDM~\cite{liu2023audioldm}, we do not perform additional model finetuning on AudioCaps for model trained with the full-scale datasets. Models with the suffix \textit{Large} indicate larger-sized model variants, such as \vModelName\textit{-Full-Large}. 




\subsection{Text-to-Audio Generation}
\noindent
We compare the performance of our proposed model with several state-of-the-art systems, including AudioGen-Large~\cite{kreuk2022audiogen}, Make-an-Audio~\cite{huang2023make-an-audio}, AudioLDM~\cite{liu2023audioldm}, Make-an-Audio 2~\cite{huang2023make-an-audio-2}, and TANGO~\cite{ghosal2023text-tango}. To generate the samples for subjective evaluation, we adopt AudioLDM-M, an AudioLDM with $652$M parameters, from HuggingFace\footnote{\url{https://huggingface.co/spaces/haoheliu/audioldm-text-to-audio-generation}} and run with $100$ reverse diffusion steps. 
The result of Make-an-Audio 2 is provided by the author~\cite{huang2023make-an-audio-2}. We use the pre-trained TANGO model open-sourced on GitHub\footnote{\url{https://github.com/declare-lab/tango}} to reproduce their result. 

As shown in Table~\ref{tab: main-audiocaps_sota-comparison}, our proposed~\vModelName\textit{-AC} significantly outperforms the previous systems across all three objective metrics. The previous best-performing system, TANGO, achieves a CLAP score of $17.6$, while our proposed system surpasses it with a substantially higher CLAP score of $24.9$.~\vModelName\textit{-Large} also attains the best KL divergence score of $0.98$, considerably improving upon the previous SoTA of $1.27$. 
For the FAD score, our model reaches $1.42$, establishing a new SoTA for text-to-audio generation. Our subjective evaluation results are mostly consistent with the objective metrics, confirming the effectiveness of \vModelName\textit{-AC}, which achieves an OVL of $3.88$ and a REL of $3.90$, surpassing AudioLDM and the previous SoTA TANGO by a significant margin. The difference between \vModelName\textit{-AC} and the GroundTruth, which are real audios from the AudioCaps dataset~\cite{kim2019audiocaps}, is merely $0.16$ and $0.18$ for OVL and REL, respectively, demonstrating the strong performance of our proposed system. 
The AudioLDM-M we used is not finetuned on the AudioCaps dataset, which may explain its degraded performance compared with the metric score reported in~\cite{liu2023audioldm}. \textcolor{black}{We observe the trend of overfitting during~\vModelName~training on the AudioCaps training set, due to the limited dataset size. To address this issue we measure the FAD score on the AudioCaps validation set every five epochs and treat the checkpoint before the FAD result shows degradations as the final model.}

\begin{figure}
    \centering
    \includegraphics[width=1.0\linewidth]{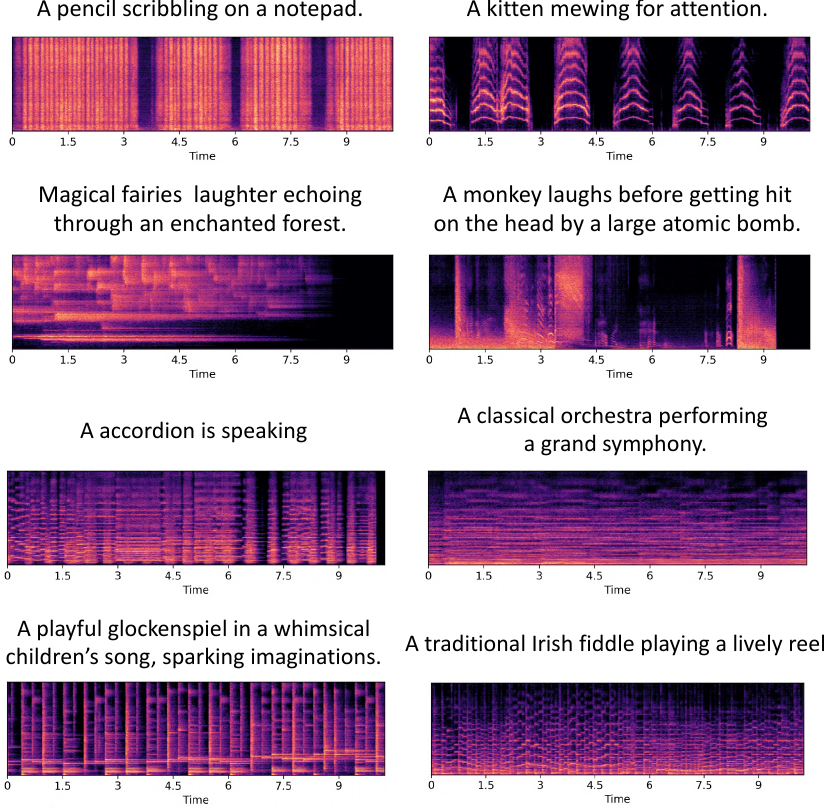}
    \caption{Examples for text-to-audio generation.}
    \label{fig: example-tta-ttm}
\end{figure}

To investigate the scalability of \vModelName\textcolor{black}{\textit{-AC-Large}}~in terms of model size and dataset scale, we further trained \vModelName~on a much larger dataset containing 29,510 hours of data using two different model sizes. \textcolor{black}{To avoid the overfitting issue and potentially misleading objective metrics result, the model trained with a larger dataset, including~\vModelName\textit{-Full} and~\vModelName\textit{-Full-Large} are not fine-tuned on the AudioCaps training set, as performed by previous works~\cite{liu2023audioldm}.}
\textcolor{black}{As shown in Table~\ref{tab: main-audiocaps_sota-comparison},} the FAD score generally shows improvement after scaling up the model size, while the KL divergence and CLAP scores do not exhibit clear improvements, indicating that scaling the model size might be more beneficial for enhancing audio quality than audio-text relations.
Despite the significant increase in training data, we did not observe significant improvements in the objective evaluation metrics. On the contrary, all three metrics showed degraded performance after training on more data. This is potentially because our test set has a limited distribution, while the large-scale training data covers a much wider distribution. The mismatch between the training and test data distributions results in poorer objective scores.

Nevertheless, when compared with the AudioLDM-M~(FAD $4.53$) in Table~\ref{tab: main-audiocaps_sota-comparison}, which is also a large-scale pre-trained text-to-audio model without finetuning on AudioCaps, \vModelName~with full-scale training data achieves significantly better performance~(FAD $1.42\sim$$2.13$), showing a substantial improvement over AudioLDM-M.

\subsection{Text-to-Music Generation} 

\begin{table}[htbp]
\centering
\scriptsize
\caption{Performance Comparison on the MusicCaps Evaluation Set. The superscript $^\star$ indicates results reproduced using publicly available implementations. The open-source version of MusicGen-Medium excludes vocal sounds, resulting in slightly inferior performance compared to the original report~\cite{copet2023simple-musicgen}. \textcolor{black}{GT-AudioMAE denote directly applying the ground truth ``Language of Audio'' $Y$, to the function $\mathcal{G}$ for audio generation, as detailed in Section~\ref{sec: overview}}. All generated audio clips were resampled to $16$kHz prior to evaluation.}
\begin{tabular}{lcccccc}
\toprule
Model               & FAD$\downarrow$   & KL$\downarrow$                           & CLAP$\uparrow$ & OVL$\uparrow$ & REL$\uparrow$ \\
\midrule
GroundTruth               & -     & -                            & $0.253$      &  $3.82$ &  $4.26$  \\
\textcolor{black}{GT-AudioMAE}               & $2.18$     & $0.27$                            & $0.257$      &  $3.59$ &  $3.92$  \\
Riffusion                & $14.80$ & $2.06$ & $0.190$       & -   & -   \\
Mousai               & $7.50$  & $1.59$                         & -          & -   & -   \\
MeLoDy             & $5.41$  & -                            & -          & -   & -   \\
MusicLM             & $4.00$  & -                            & -          & -   & -   \\
MusicGen-Medium    & $3.4$  &  $1.23$                        &     $\mathbf{0.320}$   &  -   &  -   \\
MusicGen-Medium$^{\star}$    & $4.89$  &  $1.35$                        &     $0.291$   &  $3.37$   &  $3.38$   \\
AudioLDM-M$^{\star}$             & $3.20$  & $1.29$                         & \textcolor{gray!40}{$0.360$}       &  $3.03$   &  $3.25$   \\

\midrule
\vModelName\textit{-MSD}    &    $4.47$    &  $1.32$     & $0.294$                        &  $\mathbf{3.41}$   & $3.30$ \\
\vModelName\textit{-Full}                & $\mathbf{3.13}$  & $\mathbf{1.20}$                         &   $0.301$     &  $3.34$   &  $\mathbf{3.54}$   \\
\bottomrule
\end{tabular}

\label{tab: comparison-music-generation}
\end{table}

In this section, we compare our proposed model with other text-to-music generation models, including MusicGen~\cite{copet2023simple-musicgen}, MusicLM~\cite{agostinelli2023musiclm}, MeLoDy~\cite{lam2023efficient-melody}, Mousai~\cite{schneider2023mo}, AudioLDM~\cite{liu2023audioldm}, and Riffusion~\cite{riffusion}. The output of AudioLDM is obtained in the same way as Table~\ref{tab: main-audiocaps_sota-comparison}. MusicGen is reproduced using the official Github repository\footnote{\url{https://github.com/facebookresearch/audiocraft}}. 

As shown in Table~\ref{tab: comparison-music-generation}, our proposed method significantly outperforms these strong baselines. For instance, ~\vModelName\textit{-Full} outperforms MusicGen by $36\%$, $11\%$, and $3.4\%$ on FAD, KL and CLAP scores, respectively. The~\vModelName\textit{-MSD} model, which is only trained on music data, does not achieve better performance on objective metrics than the more general~\vModelName\textit{-Full}. This result suggests that learning audio generation from a general perspective can benefit the performance in specialised domains as well, demonstrating the advantages of our proposed general framework. The general model~\vModelName\textit{-Full} achieves a significantly higher $3.54$ REL score than the other systems, indicating better textual understanding ability. 
The AudioLDM-M model achieves a significantly higher CLAP score than the remaining systems, \textcolor{black}{which may stem from being directly conditioned by the same CLAP model during training. Therefore, the CLAP score value of AudioLDM-M in Table~\ref{tab: comparison-music-generation} is only provided for reference and may not reflect the true performance of the model, as also indicated by the subjective evaluation score.}
The high performance of AudioLDM may also stem from the diversity of audio training data, which also includes music and sound effects, which further supports the benefits of training a general-purpose model. However, the subjective evaluation in Table~\ref{tab: comparison-music-generation} indicates that the subjective performance of AudioLDM-M is not as good as suggested by the objective metrics. \textcolor{black}{Since MeLoDy and MusicLM are not open-sourced, some of their objective and subjective metrics scores are not available for comparison. Due to the substantially lower objective scores, Riffusion and Mousai are not included in our subjective evaluation against other baseline models.}

\subsection{Text-to-Speech Generation} 

\noindent
We compare our proposed model with the widely-adopted FastSpeech2\footnote{\url{https://huggingface.co/facebook/fastspeech2-en-ljspeech}} model on the LJSpeech test set. To study the upper bound of our system, we add a setting called GT-AudioMAE that utilizes the ground truth LOA $Y$ to the function $\mathcal{G}$ for audio generations. Our proposed \vModelName\textit{-LJS} is trained on the LJSpeech training split. To further explore the potential of our system, we pre-train the GPT-2 model in function $\mathcal{M}$ on the GigaSpeech dataset before finetuning on LJSpeech. This version is denoted as \vModelName\textit{-LJS-Pretrained}. 

\begin{table}[htbp]
\centering
\small
\caption{Text-to-speech performance evaluated on the LJSpeech test set. }
\begin{tabular}{lc}
\toprule
     Model               & Mean Opinion Score$\uparrow$ \\
\midrule
GroundTruth         &  $4.63 \pm 0.08$   \\
GT-AudioMAE         &   $4.14 \pm 0.13$\\
FastSpeech2        &   $3.78 \pm 0.15$  \\
\midrule
\vModelName\textit{-LJS}            &   $3.65 \pm 0.21$  \\
\vModelName\textit{-LJS-Pretrained} &  $4.00 \pm 0.13$  \\
\bottomrule
\end{tabular}
\label{tab: text-to-speech}
\end{table}

As shown in Table~\ref{tab: text-to-speech}, with the pre-trained GPT-2 model, \vModelName\textit{-LJS-Pretrained} achieves a MOS of $4.00$, significantly outperforming FastSpeech2. 
Our subjective evaluation shows \vModelName\textit{-LJS-Pretrained} exhibits greater fluctuations in emotion, punctuation, and tone. This demonstrates the benefits of pretraining on diverse datasets like GigaSpeech before finetuning on smaller corpora.
Without pretraining, our proposed model still achieves a competitive MOS~(Mean Opinion Score) of $3.65$, which is comparable with the $3.78$ MOS of our baseline FastSpeech2. 

\subsection{Ablation Studies}

\begin{table}[htbp]
\centering
\scriptsize
\caption{Ablation Studies on the AudioCaps dataset.}
\begin{tabular}{lccc}
\toprule
    Setting            & FAD$\downarrow$ & KL$\downarrow$   & CLAP$\uparrow$  \\
\midrule
\vModelName    &  $1.67$   &   $\mathbf{1.01}$   &   $0.249$    \\
~~~~\textit{a. ~~w/o Joint finetuning}     &  $2.24$   &   $1.07$   &  $0.234$     \\
~~~~\textit{b. ~~w/o CLAP embedding~(GPT)}     &  $2.48$   &  $1.07$    &   $0.245$    \\
~~~~\textit{c. ~~w/o FLAN-T5 embedding~(GPT)}  &  $2.73$   &   $1.05$   &  $\mathbf{0.250}$     \\
~~~~\textit{d. ~~w/o FLAN-T5 crossattn~(T-UNet)}  & $\mathbf{1.38}$ & $1.30$ & $0.211$ \\
~~~~\textit{\textcolor{black}{e. ~~w/o CLAP and FLAN-T5~(GPT)}}  &  $2.11$   &   $1.06$   &  $0.213$     \\
\bottomrule
\end{tabular}
\label{tab: ablation}
\end{table}

\noindent
In order to validate our design choices of~\vModelName, we conducted a series of ablation studies on the text-to-audio generation task on the AudioCaps dataset. The results are shown in Table~\ref{tab: ablation}. When the joint finetuning process between the GPT-2 model and the latent diffusion model was disabled~(\textit{a}), thereby only optimizing them separately, all three evaluation metrics exhibited a marked deterioration, suggesting joint fine-tuning is helpful for the GPT-2 model to better cooperate with the LDM model. The GPT-2 model accepts inputs from both the CLAP and FLAN-T5 modules for text-to-audio generation. The removal of either module resulted in a degradation of the evaluation metrics~(\textit{b-c}). However, the CLAP score was improved when only the CLAP module was used as an input~(\textit{c}). This improvement is likely due to the conditioning directly matching the evaluation metric. The removal of the cross-attention mechanism in the T-UNet model~(\textit{d}), which accepts the FLAN-T5 embeddings, led to a significant degradation in both the KL divergence and CLAP~scores. However, it improved the FAD score, from $1.67$ to $1.38$. These results indicate that while AudioMAE conditioning alone can achieve better FAD, the use of FLAN-T5 conditioning provides additional language semantic information that assists the learning of the audio and text relationships. \textcolor{black}{Besides, we study the effect of removing both CLAP and FLAN-T5 representations and directly use text as input to the GPT-2 model to predict LOA~\textit{(e)}. The experimental result shows that our model in this setting maintains competitive performance with an FAD of $2.11$ and KL of $1.06$. However, the CLAP score exhibits a noticeable degradation, which indicates that the CLAP and FLAN-T5 representations can potentially improve the relationship between the text and the generated audio.}

\section{Conclusion and Future Works}
\noindent
In this paper, we have presented~\vModelName~for audio generation, achieving state-of-the-art or comparative performance on text-to-audio, text-to-music, and text-to-speech generation tasks. 
As a universal audio representation, the language of audio~(LOA) enables self-supervised pre-training of the latent diffusion model, providing a robust foundation for the audio generation task.
We further demonstrate the versatility of our proposed method by performing audio in-context learning. \vModelName~opens doors for future works on audio generation from a unified perspective. Future work includes enabling the multi-task learning of the GPT-2 model to generate audio, music, and speech simultaneously with a single model. \textcolor{black}{Additionally, we plan to investigate more effective representations for the "language of audio" by exploring the integration of other audio self-supervised models, such as HuBERT~\cite{hsu2021hubert} and wav2vec~\cite{baevski2020wav2vec}, into our system.}

\section*{Acknowledgments}
\noindent
This research was partly supported by the British Broadcasting Corporation Research and Development~(BBC R\&D), Engineering and Physical Sciences Research Council (EPSRC) Grant EP/T019751/1 ``AI for Sound'', and a PhD scholarship from the Centre for Vision, Speech and Signal Processing (CVSSP), Faculty of Engineering and Physical Science (FEPS), University of Surrey. For the purpose of open access, the authors have applied a Creative Commons Attribution (CC BY) license to any Author Accepted Manuscript version arising. \textcolor{black}{The authors wish to thank the associate editor and the reviewers for their helpful comments to further improve this work.}

\bibliography{reference}
\bibliographystyle{IEEEtran}

\vfill

\end{document}